\documentstyle[preprint,prb,aps]{revtex}

\begin{document}
\draft
\preprint{}
\title{Marshall sign and the doped $t-J$ model}
\author{ Z.Y. Weng, D.N. Sheng, and Y.C. Chen* }
\address{Texas Center for Superconductivity and Department of Physics\\
University of Houston, Houston, TX 77204-5506 }
\maketitle
\begin{abstract}
Marshall sign as a sole source of sign problem hidden in an antiferromagnet is
explored under doping. By tracking the Marshall sign, a zero spectral weight
$Z$ is revealed in the doped
antiferromagnetic system. $Z=0$ is
caused by a phase string induced by the ``bare'' hole. By eliminating
such a phase string through nonlocal transformations, a non-perturbative scheme
is obtained. It is argued that this formalism provides a unique way
to get access to the real ground state of the doped $t-J$ model for both one-
and two-dimensions.

\end{abstract}

\vspace{0.4in}

\narrowtext

According to Marshall,\cite{marshall} the ground-state wavefunction of the
Heisenberg Hamiltonian for a bipartite lattice is real and satisfies a
sign rule. This sign rule can be uniquely determined by requiring that a flip
of any two antiparallel spins at nearest-neighboring
sites always involves a sign change in the wave function: $\uparrow\downarrow
\rightarrow (-1)\downarrow\uparrow$. Such a Marshall sign has played a
crucial role in the success
of two types of approximate approaches: the resonant-valence-bond (RVB)-type
variational
wavefunction\cite{liang} proposed by Liang, Doucot and Anderson, which gives
one of the lowest energy bound (-0.3344 J/per bond), and the Schwinger-boson
mean-field state\cite{AA}  where the Marshall sign is incorporated in the order
parameter $<\sum_{\sigma}\sigma \bar{b}_{i\sigma}\bar{b}_{j-\sigma}>$ ($i$ and
$j$ are the nearest-neighboring sites).

Difficulty arises when one tries to dope holes into this antiferromagnet.
Doped holes are expected to mess up with the Marshall sign, and the latter
becomes a crucial source of sign problem hidden in the spin background. Doping
effect is described by the well-known $t-J$ model as follows
\begin{eqnarray}\label{e0}
H_{t-J}& =&J\sum_{<ij>}\left({\bf S}_i\cdot {\bf S}_j-\frac {n_in_j}
4 \right) -t \sum_{<ij>\sigma} (c_{i\sigma}^{\dagger} c_{j\sigma}+H.c. )
\nonumber \\
       & \equiv &H_J+H_t.
\end{eqnarray}
in which the Hilbert space is restricted by the no-double-occupancy constraint
$\sum_{\sigma}c^{\dagger}_{i\sigma}c_{i\sigma}\leq 1$. At the zero-doping
limit,
$H_J$ recovers the antiferromagnetic Heisenberg model.  The $t-J$ model has
been
intensively studied in recent years due to its widely perceived connection with
the high-$T_c$
problem. But very limited understandings about this model have been
achieved in two dimensions (2D) because of its strongly-correlated nature.

In this paper, we shall clarify the non-perturbative characteristics
of the model in terms of the Marshall sign. A new representation is proposed in
which the sign problem can be totally
resolved in one dimension (1D), while partially eliminated in 2D with the
residual phase problem tracked exactly through some topological phases.
This non-perturbative approach provides an accurate way to understand
long-wavelength physics of the $t-J$ model for both 1D and 2D.

The importance of keeping the Marshall sign in the undoped Heisenberg model may
be understood as follows: under a spin basis $|\phi>$ with the aforementioned
Marshall sign included, the matrix element of
the Heisenberg Hamiltonian becomes negative-definite, i.e., $<\phi'|H|\phi>\leq
 0$. Then the ground state $|\psi_0>=\sum_{\phi}c_{\phi}|\phi>$ will always
have
real, positive coefficient (or wave-function) $c_{\phi}$. In other words, the
Marshall sign is indeed the sole source of signs in the ground state, and there
is no more sign problem in this new representation. Loosely speaking, physical
properties will be less sensitive to various approximations here, which is the
basic reason for the success of those approaches\cite{liang,AA} mentioned
before.

The Marshall sign described above can be easily built into the $S^z$-spin
representation even in the presence of holes.\cite{weng1} The bipartite lattice
can be divided into even ($A$)
and odd ($B$) sublattices. For each down spin at $A$ site or up spin at $B$
site, one may assign an extra phase $i$ to the basis. In this way, a flip of
two nearest-neighboring spins will always involve a sign change (i.e., the
Marshall sign): $\uparrow\downarrow \rightarrow (i^2)\downarrow\uparrow = (-1)
\downarrow\uparrow$. Of course, this is not a unique way to incorporate the
Marshall sign in the spin basis, but it will be  quite a useful bookkeeping
once holes are introduced into the system. Generally, the spin-hole
basis may be defined as
\begin{equation}\label{e1}
|\phi>= i^{N_A^{\downarrow}+N_B^{\uparrow}} |\uparrow ...\downarrow\uparrow
\circ ...\downarrow>  ,
\end{equation}
with $N_A^{\downarrow}$ and $N_B^{\uparrow}$ as total numbers of down and up
spins
at A- and B-sublattices, respectively. It is straightforward to verify that the
matrix element
\begin{equation}\label{e2}
<\phi'|H_J|\phi> \leq 0,
\end{equation}
under this new basis even in the presence of holes.

However, once the holes are allowed to move around, they will cause serious
sign problem. For the sake of clarity, we first consider a single hole problem,
where the statistics problem among holes is absent.  Suppose the hole initially
sitting  at site $n$ hops onto a nearest-neighboring site $m$. The
corresponding  matrix element can be found to be
\begin{equation}\label{e3}
<\phi_{(m)}|H_t|\phi_{(n)}>= -t \times (\pm i) ,
\end{equation}
where the subscripts $(n)$ and $(m)$ denote hole's sites in the basis (2), and
$(\pm i)$ is determined by the original spin state $\sigma_m=\pm 1$ at site
$m$:
\begin{equation}\label{e4}
(\pm i) \equiv (-1)^m e^{i\frac {\pi} 2 \sigma_m},
\end{equation}
where $(-1)^m$ is the staggered factor: $(-1)^A=+1$ and $(-1)^B= -1$. Thus a
doped hole moving around will always leave a trace of phases
(phase string) $(\pm i)\times (\pm  i)\times ...$  behind it. This phase string
cannot be eliminated through spin-flip process described by (\ref{e2}) since
the latter does not produce extra ``signs''.  It implies that each doped
hole always creates a phase string in the spin background which is not
repairable at low energy, and thus it will determine the long-distance and
long-time behaviors of the hole, as to be discussed below. In the conventional
approximations, this important effect has been overlooked because the Marshall
sign is not properly tracked in the doping problem.

A bare doped hole is described by $c_{i\sigma}|\psi_0>$. One may follow the
evolution of the doped hole by studying its propagator
$G_1(j,i; E)=<\psi_0|c_{j\sigma}^{\dagger}
(E-H)^{-1}c_{i\sigma}|\psi_0>$. By using the expansion
\begin{equation}
G_1=<\psi_0|c_{j\sigma}^{\dagger}\left( G_0^J+G_0^JH_tG_0^J+... \right)
c_{i\sigma}|\psi_0>,
\end{equation}
where $G^J_0=(E-H_J)^{-1}$, one finds the contribution to each path, connecting
$i$ and $j$, is weighted by a corresponding  phase string $(\pm i)\times (\pm
i)
\times ...=\Pi_{m}i(-1)^m\sigma_m$ in terms of (\ref{e3}) and (\ref{e4}). The
rest factors are found to be
sign-definite since each term $<\phi'_{(m)}|G_0^J(E)|
\phi_{(m)}>$ ($m$ is a hole-site on the path) is always negative, shown by
using the expansion $G_0^J=1/E\sum_k (H_J/E)^k$ and the condition (\ref{e2}).
(The expansion series is converged at least when $E$ is less than the lower
energy bound $E_G^0$ of $H_J$ (with a hole fixed
at site $m$)). Due to the accumulated effect, the phase string $(\pm i)\times
(\pm i)\times ... $ can be straightforwardly shown to lead to a vanishing
contribution for each given path beyond  the spin-fluctuational correlation
length, after being averaged over the various spin configurations.\cite{weng4}
Thus the bare hole will lose its coherence as it cannot travel over a large
distance. In turn, it means a vanishing spectral weight $Z(E)$ (e.g., $Z(E_G)=
|<\psi|c_{k\sigma}|\psi_0>|^2$, etc., where $|\psi>$ and $E_G$ are the
ground state and its energy, respectively) at least when $E\rightarrow
E_G$.\cite{remark} A more rigorous demonstration of
$Z=0$ for a one-hole doping problem will be given
in a separate paper.\cite{weng4} In the above discussion, the
Hamiltonian properties (\ref{e2})-(\ref{e4}) is crucial
in leading to $Z=0$ for the one-hole case. It is easy to see that even at a
sufficiently small doping, where the additional sign problem due to  fermionic
statistics among doped holes are not important, the conclusion of $Z=0$ is
still robust.

$Z=0$ means that there is no
overlap between $c_{i\sigma}|\psi_0>$ and the ground state (and low-lying
excitation states). Each doped hole will have to induce a {\it global}
adjustment of the spin background in order to reach the ground state. So one
may not be able to get access to the true ground state {\it perturbatively} by
starting from
$c_{i\sigma}|\psi_0>$. But $Z=0$ itself does not tell how the non-perturbative
approach should be pursued. Thus we have to go back to the original source
which causes $Z=0$. In the present case, it is due to the phase string
introduced by hole, which cannot be ``repaired'' by spin-flip process. One may
regard this as a new sign problem associated with the hopping matrix element
(\ref{e3}) under the spin-hole basis (\ref{e1}), where hole is treated as a
``bare'' one. It is then natural to ask if such a sign problem can be
eliminated through some non-perturbative transformation.

{\it One-dimensional case.}\hspace{0.4in} For a single hole case, one may
define the
following modified spin-hole basis in terms of (\ref{e1}):
\begin{equation}\label{e7}
|\tilde{\phi}_{(n)}>=e^{i\Theta_n}|\phi_{(n)}>,
\end{equation}
where $n$ specifies the hole site, and
\begin{equation}\label{e8}
\Theta_n=\frac {\pi} 4 [1+ (-1)^n] + \frac 1 2 \sum_l{}' \theta_n(l)
(\sigma_l -1).
\end{equation}
In (\ref{e8}), the summation runs over all the spin sites on the chain and
$\sigma_l=\pm 1$ describes the spin state at $l$ site.
And $\theta_n(l)$ may be defined as
\begin{equation}\label{e9}
\theta_n(l)= \mbox {Im ln} (z_n-z_l),
\end{equation}
where $z_n=x_n+iy_n$ is a 2D complex coordinate and the 1D chain is sitting on
the real axis. Then one has $\theta_n(l)-\theta_l(n)= \pm \pi$. It is
straightforward to verify that $<\tilde{\phi}_{(m)}|H_t|\tilde{\phi}_{(n)}>
=-t$, while the matrix element for $H_J$ remains the same as (\ref{e2}). In the
many-hole case, $\Theta_n$ in (\ref{e7}) should be replaced by a total
phase-shift $\Theta=\sum_n^{''} \Theta_n$ (the summation is over the hole
sites) in additional to a fermionic-statistics factor
$e^{-i\sum^{''}_{n<n'}\theta_n(n')}$, and the conclusions remain the same.

Thus, the sign problem in 1D can be completely eliminated in the new
representation,
and the exact ground state expanded in terms of this basis always has
real, positive-definite coefficient. According to (\ref{e7}), then, each hole
induces a nonlocal phase-shift $\Theta_n$ in the true ground state. It
represents a
non-perturbative change of the system and is consistent with $Z=0$ discussed
before. We note that the phase-shift idea in 1D was first proposed by
Anderson,\cite{anderson} and here its accurate form is simply obtained by
tracking the Marshall sign.

It is interesting to express the original electron operator in this new
representation.
A bare hole created by $c_{i\sigma}$ will lead to a phase-shift
$e^{-i\Theta_i}$ in the new representation. Furthermore, the original spin
$\sigma$ at $i$ site has a phase $e^{i\pi/4[1-\sigma(-1)^i]}$ (Eq.(\ref{e1}))
and a contribution to other holes $e^{i\sum_{m\neq
i}''1/2\theta_m(i)(\sigma -1)}$. One may introduce a holon creation operator
$h_i^{\dagger}$ and a spinon annihilation operator $b_{i\sigma}$ to keep the
track of charge and spin in the new representation (both are bosonic
operators), then the electron annihilation operator can be determined as
(up to a global constant phase)
\begin{equation}\label{e10}
c_{i\sigma}= h_i^{\dagger}b_{i\sigma}\left[e^{i\frac 1 2 \sum_{l\neq i}
\theta_i(l)\left(\sigma n_l^h-\sum_{\alpha}\alpha n_{l\alpha}^b +1\right)}
(-\sigma)^i\right].
\end{equation}
$n_l^h$ and $n_{l\alpha}^b$ in (\ref{e10}) are holon and spinon number
operators, satisfying the no-double-occupancy constraint $n^h_l+\sum_{\sigma}
n^b_{l\sigma} =1$.  It is easy to verify that $\{c_{i\sigma},
c_{j\sigma}^{\dagger}\}=\delta_{i,j}$ and $\{c_{i\sigma},
c_{j\sigma}\}=0$, but for opposite spins, $[c_{i\sigma},
c_{j-\sigma}^{\dagger}]=0$. The latter result may be a little bit surprising
but is physically correct. Even though the $t-J$ model is usually
formulated in terms of $c_{i\sigma}$, which satisfies anti-commutation
relations
for both spins, it is easy to show that the commutation relations for
electrons
with opposite spins are not important and one may always assign either
commutation or anti-commutation relations to them without changing the
physical consequences.\cite{yang}

{\it Two-dimensional case.} \hspace{0.4in} In 2D, a bare hole moving through
any closed path back to its origin will always leave a phase string if the path
is
not a retraceable one. It suggests that the phase problem
in 2D become quite different from the 1D case. In a one-hole problem,
one may still use the  transformations (\ref{e7})-(\ref{e9}) to eliminate  the
phase string induced by the hole. Actually, this
procedure is the {\it only} way to eliminate the phase strings on all
paths: the spins have to know the hole's position nonlocally to adjust
themselves. But in 2D one finds
\begin{equation}\label{e11}
<\tilde{\phi}_{(m)}|H_t|\tilde{\phi}_{(n)}>_M=-t e^{iA_{nm}^f},
\end{equation}
where a phase $A_{nm}^f$ is contributed by all the spins on the lattice
other than $n$ and $m$ sites:
\begin{equation}\label{e12}
A_{nm}^f=\frac 1 2 \sum_{l\neq n,m} (\theta_n(l)-\theta_m(l)) \left[\sum_{
\alpha}\alpha n_{l\alpha}^b -1\right].
\end{equation}
$A_{nm}^f$ (it vanishes in 1D) satisfies the following topological condition
\begin{equation}\label{e13}
\!\!\!\!\!\!\sum_C A_{nm}^f= \pi \sum_{l\in C}\left(\sum_{\alpha}\alpha n_{l
\alpha}^b-1\right) ,
\end{equation}
where the l.h.s. sum is over an arbitrary closed path $C$ on the lattice, while
the r.h.s. sum is over all the sites included by the path $C$.
Hence, instead of leaving a phase string, a hole in the new representation now
sees fictitious
fluxes enclosed after moving through a closed path. These fluxes are composed
of fictitious
$\pi$-flux quanta (pointing at $\hat{z}$-direction) bound to spins
on the top of a uniform lattice $\pi$-flux.
So $A_{nm}^f$ cannot be simply gauged away in 2D.  However,
critically different from the afore-discussed singular phase string, those
flux quanta
generally would not prevent the hole to travel across the whole system or, in
other words,
one can have quasiparticle-like description of the hole (holon) in this new
representation. A similar topological phase can be found in the matrix element
of  $H_J$ (see below).

Generalization to the many-hole case is
also similar to 1D. Since we have already introduced holon and spinon
operators
$h_i$ and $b_{i\sigma}$, it is more transparent to write down $H_{t-J}$ in
the new representation in operator formalism:\cite{remark2}
\begin{equation}\label{e14}
H_J=-\frac J 2 \sum _{<ij > \sigma\sigma' } \left( e^{i\sigma A_{ij}^h}b_{i
\sigma }^+
b_{j\sigma }\right)\left( e^{i\sigma' A_{ji}^h}b_{j \sigma' }^+b_{i\sigma'}
\right)+ \frac J 2 \sum_{<ij>}(1-n^h_i)n^h_j ,
\end{equation}
\begin{equation}\label{e15}
H_t=-t\sum_{<ij>}\left(e^{iA_{ij}^f} h_i^+h_j\right)\left( e^{i\sigma A_{ji}^h}
b_{j \sigma}^+b_{i\sigma }\right) .
\end{equation}
$A_{ij}^h$ in (\ref{e14}) and (\ref{e15}) satisfies
the following topological condition
\begin{equation}\label{e16}
\!\!\!\!\!\!\sum_C A_{ij}^h= \pi \sum_{l\in C} n_l^h .
\end{equation}

Equations (\ref{e14}) and (\ref{e15}), together with (\ref{e10}), represent an
exact reformulation of the $t-J$ model. More importantly, the ground state
and low-lying
states may become  ``perturbatively'' accessible in this new representation in
terms of new ``particles''
described by $h$ and $b_{\sigma}$. In 1D, the Hamiltonians
({\ref{e14}) and (\ref{e15}) becomes ``trivial'' without the presence of sign
problem ($A^f_{ij}=A_{ij}^h=0$, and  both $h$ and $b_{\sigma}$ are bosonic
operators). All the
non-trivial information about the Luttinger-liquid behaviors is now included as
``phase-shift'' in the $c_{i\sigma}$ operator expression (\ref{e10}).
Crucial asymptotic correlations can be correctly obtained\cite{weng1,weng2} in
this new framework even within conventional ``mean-field-type'' treatment of
(\ref{e14}) and (\ref{e15}).  Here $h$ and $b_{\sigma}$ naturally describes the
quasipartice-like ($Z_h$,$Z_s$ $\neq 0$) holon and spinon excitations, and thus
the decomposition (\ref{e10}) represents a ``correct'' spin-charge separation.

Finite $A^f_{ij}$ and $A^h_{ij}$ in 2D mean that holes and spins have to
``feel'' each other nonlocally in order to eliminate singular phase strings
created by the ``bare'' holes. Or {\it vice versa},  those phase strings'
superposition in a large-distance, long-time scale will lead to a delicate
topological effect described by $A^f_{ij}$ and $A^h_{ij}$. Thus charge and spin
degrees of freedom are intrinsically coupled together in 2D. For example,
$A^h_{ij}$ reflects doping effect on spin background. At the zero doping limit,
$A^h_{ij}$ vanishes in (\ref{e14}), and $H_J$ reduces back to a similar form as
in the Schwinger
boson representation\cite{AA} with the Marshall sign being absorbed. In the
latter formalism,  a mean-field treatment can yield a good approximation.\cite
{AA} But in the Schwinger-boson approach, the Marshall sign is usually
incorporated in the order parameter $<\sum_{\sigma}\sigma \bar{b}_{i\sigma}
\bar{b}_{j -\sigma}>$. When new mean-field order-parameters are
introduced\cite{spiral} at finite doping,
there is no systematical way to keep the track of sign. By contrast, in the
present representation, the whole essential phase at finite doping is
exactly tracked
by $A^h_{ij}$ and $A^f_{ij}$ in (\ref{e14}) and (\ref{e15}). Then any
approximations leaving $A^h_{ij}$ and $A^f_{ij}$ intact should not dramatically
affect the phase problem which is presumably crucial for the long-wavelength,
low-energy physics.

In fact, the same $A^h_{ij}$ and $A^f_{ij}$ as well as the decomposition (10)
have been already identified from
a different approach\cite{weng2} recently. $A^h_{ij}$ in 2D has been shown
there to lead to
the deconfinement of spinons in (\ref{e14}) and of spinon-holon in (\ref{e15}),
and  introduce a finite spin-spin
correlation length in a fashion of $1/\sqrt{\delta}$ ($\delta$ is the doping
concentration). An anomalous transport phenomenon induced by $A_{ij}^f$ has
been also studied. The gapped transverse gauge fluctuation ensures the
accuracy of the topological phases $A^h_{ij}$ and $A^f_{ij}$ in
long-wavelength, low-energy regime. Therefore, like in 1D, one finds 2D
spin-charge separation in this formalism, and $h$ and $b_{\sigma}$ properly
describe the elementary charge and spin excitations (holon and spinon),
respectively. Strong experimental features in association with the
normal state of cuprates have been found in both spin and charge channels in
this approach.\cite{weng2}

Finally,  we would like to briefly discuss the slave-boson formalism
$c_{i\sigma}=h^{\dagger}_if_{i\sigma}$. If one writes $S_i^{+}S_j^{-}=
-(f_{i\uparrow}^{\dagger}f_{j\uparrow}f^{\dagger}_{j\downarrow}f_{i\downarrow})
$,
it seems that the Marshall sign is automatically preserved here. Nonetheless,
extra phase problem is introduced by the fermionic operator $f_{i\sigma}$. It
is reflected in, for instance, $<f_{i\sigma}^{\dagger}f_{j\sigma}>=
\sum_ke^{ik\cdot(x_j-x_i)}<f^{\dagger}_{k\sigma}f_{k\sigma}>$ where a lot of
$k$'s must be involved due to the Pauli-principal. Recall that a free
fermion excitation in the bosonic representation would be described as a
vortex, and {\it vice versa}. Then from the  present point of review, at least
close to the half-filling one cannot get access
to the true ground-state ``perturbatively'' by starting with this formalism.
It is also noted that at large doping, $A^h$ in the present scheme can even
split, in terms of the no-double-occupancy constraint, such that to become the
statistics-transmutation phases which can turn spinons into fermionic ones to
recover the usual Fermi-liquid behavior.

In conclusion, by carefully examining the Marshall sign, we have shown that a
doped hole will induce a string-like phase defect in the spin background.
This phase string cannot be removed by low-lying spin fluctuations, and thus
causes a vanishing quasiparticle spectral weight $Z$. A nonlocal transformation
is found to eliminate such a phase string in both 1D and 2D. As a result, sign
problem is totally resolved in 1D, while the residual sign
problem in 2D is kept tracked through some topological phases. This is a
non-perturbative scheme with regard to the original
electron description where a global phase shift  is present due to the
superposition of phase strings caused by doped holes. We argue that
in this new representation the ground-state and low-lying states can be {\it
perturbatively } approached, and thus
provide a unique way to systematically understand the weakly-doped $t-J$ model.
It also lends a crucial support and justification for a recent approach\cite
{weng2} based on different physical principal, which exhibits exactly the
same basic structure as in the present representation. The spin-charge
separation is identified there for both 1 \& 2D, and the magnetic and
transport anomalies are found in striking similarities with the high-T$_c$
cuprates.

\acknowledgments
The present work is supported partially by
Texas Advanced Research Program
under Grant No. 3652182 and
by Texas Centre for Superconductivity at University of Houston.

* Permanent address: Department of Physics, University of Science and
Technology
of China, Hefei, Anhui 230026, China.


\begin{references}
\bibitem{remark} An another possibility is that the effective mass of the doped
hole becomes infinite near the ground-state energy. Because of the
dispersionless of the quasiparticle spectrum in this case, a continuum of
momenta $k$'s can be involved in the ground state which can also lead to a
decay of $G_1$ at $|x_i-x_j|\rightarrow \infty$ and $E=E_G$. But this case
could be excluded in 1D and 2D as discussed in Ref.{\onlinecite{weng4}}.
\bibitem{remark2} The hard-core boson condition $b_{i\sigma}b_{i\sigma}^{
\dagger}+b^{\dagger}_{i\sigma}b_{i\sigma}=1 $ is used
in re-arranging the  $S^z$ component of $H_J$.
\end{references}
\end{document}